\documentstyle[aps,12pt]{revtex}
\widetext
\draft
\tighten

\begin{document}

\title{Reply to the Comment on ``Theorem on the proportionality of
inertial and gravitational masses in classical mechanics"}

\bigskip

\author{\bf Andrew E. Chubykalo and Stoyan J. Vlaev}

\address {Escuela de F\'{\i}sica, Universidad Aut\'onoma de Zacatecas \\
Apartado Postal C-580\, Zacatecas 98068, ZAC., M\'exico}

\date{\today}

\maketitle


\baselineskip 7mm

\begin{abstract}
     In a preceding Comment, the author declares that we claim that the
ratio of inertial mass to gravitational mass can be derived {\it ex nihilo}
and that our paper was published by mistake.  In this ``Reply" we dispute
the point of view of the author.
\end{abstract}



     In the preceding Comment [1], B. Jancovici tries to
argue why in his  opinion our proof makes no sense: ``{\it Reading the
paper, one realized that it contains irrelevant calculations ... aiming to
prove that $\eta$ is a constant for one given body: how could the ratio of
two constants be something else?"} The point is that after reading our
paper, anyone can realize that we {\it do not use} a postulate
of classical mechanics that mass of a body is a {\it constant}. In
other words, we prove that ratio $\eta$ of inertial $m_i$ to gravitational
$m_g$ is a {\it constant} without postulating that masses $m_i$ and $m_g$
are absolute constants. To be more specific, let us compare postulates
used by a generally accepted classical mechanics (GACM) with the
postulates used in our proof.

Postulates of GACM:\\
$(a_0)$ {\it any} body with non-zero inertial mass possesses also non-zero
gravitational mass;\\
$(a)$ (the Equivalence principle) the inertial mass of the body is
proportional to the gravitational mass of the same body and a constant of
the proportionality is the same for {\it all} bodies;\\
$(b)$ the masses
of bodies (inertial and gravitational) are absolute constants
(invariant);\\ $(c)$ masses obey the principle of additivity.

{\bf Note:} In GACM the claim that
\begin{equation}
{\rm from}\quad (a_0)+(b) + (c) \Rightarrow(a)
\end{equation}
is not quite obvious from a theoretical point of view: for example, let us
consider two {\it concrete} bodies with the different masses $m_1$ and
$m_2$.  Their {\it constant} masses $m_{i1},m_{i2},m_{g1},m_{g2}$,
obviously, must obey the relations
\begin{equation}
m_{i1}/m_{g1}=\eta_1\qquad{\rm and}\qquad
m_{i2}/m_{g2}=\eta_2
\end{equation}
but, generally speaking, it is not obvious that
\begin{equation}
\eta_1=\eta_2.
\end{equation}
So in GACM one considers
Eq.(3) as an experimental fact. However, after one applies our arguments
(see Eqs. (15)-(21) in [7]) one can prove the validity of the claim (a)
(or Eq.(3)).

{\bf An important remark about the postulate $(b)$:} in his famous book [2]
E.Mach convincingly shows (analyzing Newton's well-known experiment with
the ``revolving pail") that all experimentally verifiable equations of
Newtonian classical mechanics do not change if one supposes that inertial
mass of a body is not an absolute constant (invariant) and, generally
speaking, it can depend on the location of a body in space. So many
scientists (see, e.g., [3-5]), following Mach's ideas, expect that the
principle of the proportionality of $m_i$ and $m_g$ may not be valid for
classical mechanics.  That is why when constructing general
relativity, Einstein started with the Mach principle, but had to reject
it thereafter (e.g.,see [6]) because of its disagreement (as Einstein
believed) with the Equivalence principle.  And that is why a proof of the
postulate $(a)$ also in the framework of Mach's ideas is of great
interest.

Our postulates:\\
$(a_0)$ {\it any} body with non-zero inertial mass possesses also non-zero
gravitational mass;\\
$(b_1)$ the masses of bodies (inertial and gravitational) do not depend
explicitly on time but they {\it can} depend on their
location in space;\\
$(c_1)$ both inertial and gravitational masses obey the principle of
additivity.

It is obvious that our postulates are weaker than the former ones.
One can see that we do not  conserve the point $(a)$ as
a postulate,  bat what's more, our postulate $(b_1)$
sufficiently differs from the postulate $(b)$.  The validity of the
point $(a)$ from the postulates
$(a_0),(b_1),(c_1)$ also (compare with Eq.(1)) is not obvious, because
$m_i$ and $m_g$ may depend on  location {\it in different ways}. This is
so, because $m_i$ and $m_g$ have a {\it different} origin in classical
mechanics.

In our paper we proved that from our postulates $(a_0)$, $(b_1)$ and
$(c_1)$ one infers the claim $(a)$. Actually, we proved that if according
to the Mach principle, the inertial mass of a body can change from point
to point in space, then the gravitational mass of the same body {\it must}
also change by the same law, i.e. $m_i$ and $m_g$ are linear dependent
(proportional) one-to-one functions.  In  other words, we show that even
in the framework of the Mach principle the proportionality of inertial and
gravitational masses {\it must} take place.

In the last paragraph of the Comment, the author advances his  most
serious critical remark. However, at this
point he misses an implicit but very important factor:

Indeed, after one applies our arguments (see
[7], Eqs.  (6)-(11)), and provided that one neglects electromagnetic
radiation, we can formally obtain an expression
\begin{equation}
m_i={\rm
const}\;q_m,
\end{equation}
(where $q_m$ is a charge of the {\it given}
body $m$) for a  {\it concrete} body with a {\it given} inertial mass
and a {\it given} velocity. But we {\it cannot} apply the subsequent
arguments (see [7], Eqs.(15)-(21)) in order to ``prove" that {\it all}
particles have the same charge-to-mass ratio. The point is that in our
speculations we implicitly use the postulate $a_0$, namely, ``{\it any}
body with non-zero inertial mass possesses also non-zero gravitational
mass".  In  other words, after  Eq. (11) (in [7]) we {\it can} claim
that for bodies having the same mass, their $m_i$ satisfy Eq.(11)
from [7]. But we {\it cannot} claim {\it this} in the discussion
{\it concerning} ``inertial mass-charges" relation. In the latter
case one should postulate {\it that any body with non-zero inertial mass
possesses also non-zero charge}.  It is obvious that such a ``postulate"
does not follow from the experiment.

As a finishing remark, let us note that reading the Abstract of the
``Comment", a reader might come to the conclusion that our paper was
 published by mistake, after being rejected by the Editorial Board.
In fact,  the paper was published with minor modifications following
a positive  report by an anonymous referee selected by the Board.
However, we recognize that the critical remarks contained in the
``Comment" are partly justified:  we should have explicitly defined our
postulates.

\end{document}